\PassOptionsToPackage{dvipsnames}{xcolor}
\documentclass[10pt,sigconf,nonacm,screen]{acmart}

\usepackage[all]{nowidow}
\usepackage{xcolor}
\usepackage{subcaption}
\usepackage{hyperref}
\usepackage{acronym}

\usepackage[capitalise,noabbrev]{cleveref}
\crefformat{section}{\S#2#1#3}
\crefformat{subsection}{\S#2#1#3}
\crefformat{subsubsection}{\S#2#1#3}
\crefrangeformat{section}{\S#3#1#4 to~\S#5#2#6}
\crefrangeformat{subsection}{\S#3#1#4 to~\S#5#2#6}
\crefrangeformat{subsubsection}{\S#3#1#4 to~\S#5#2#6}

\definecolor{darkblue}{rgb}{0.0, 0.0, 0.55}
\newcommand{\name}{\textsc{Minos}}
\newcommand{\ef}{elysium threshold}

\newcommand{\EF}{Elysium Threshold}

\begin{document}

\author{Trever Schirmer}
\affiliation{%
    \institution{TU Berlin}
    \city{Berlin}
    \country{Germany}}
\email{ts@3s.tu-berlin.de}
\orcid{0000-0001-9277-3032}

\author{Natalie Carl}
\affiliation{%
    \institution{TU Berlin}
    \city{Berlin}
    \country{Germany}}
\email{nc@3s.tu-berlin.de}
\orcid{0009-0000-5991-9255}

\author{Nils Höller}
\affiliation{%
    \institution{TU Berlin}
    \city{Berlin}
    \country{Germany}}
\email{nih@3s.tu-berlin.de}

\author{Tobias Pfandzelter}
\affiliation{%
    \institution{TU Berlin}
    \city{Berlin}
    \country{Germany}}
\email{tp@3s.tu-berlin.de}
\orcid{0000-0002-7868-8613}

\author{David Bermbach}
\affiliation{%
    \institution{TU Berlin}
    \city{Berlin}
    \country{Germany}}
\email{db@3s.tu-berlin.de}
\orcid{0000-0002-7524-3256}

\title{\name{}: Exploiting Cloud Performance Variation with Function-as-a-Service Instance Selection}


\begin{abstract}
    Serverless Function-as-a-Service (FaaS) is a popular cloud paradigm to quickly and cheaply implement complex applications.
    Because the function instances cloud providers start to execute user code run on shared infrastructure, their performance can vary.
    From a user perspective, slower instances not only take longer to complete, but also increase cost due to the pay-per-use model of FaaS services where execution duration is billed with microsecond accuracy.
    In this paper, we present \name{}, a system to take advantage of this performance variation by intentionally terminating instances that are slow.
    Fast instances are not terminated, so that they can be re-used for subsequent invocations.
    One use case for this are data processing and machine learning workflows, which often download files as a first step, during which \name{} can run a short benchmark.
    Only if the benchmark passes, the main part of the function is actually executed.
    Otherwise, the request is re-queued and the instance crashes itself, so that the platform has to assign the request to another (potentially faster) instance.
    In our experiments, this leads to a speedup of up to 13\% in the resource intensive part of a data processing workflow, resulting in up to 4\% faster overall performance (and consequently 4\% cheaper prices).
    Longer and complex workflows lead to increased savings, as the pool of fast instances is re-used more often.
    For platforms exhibiting this behavior, users get better performance \emph{and} save money by wasting more of the platforms resources.
\end{abstract}

\maketitle

\section{Introduction}
\label{sec:introduction}

\begin{figure}
    \centering
    \includegraphics[width=\linewidth]{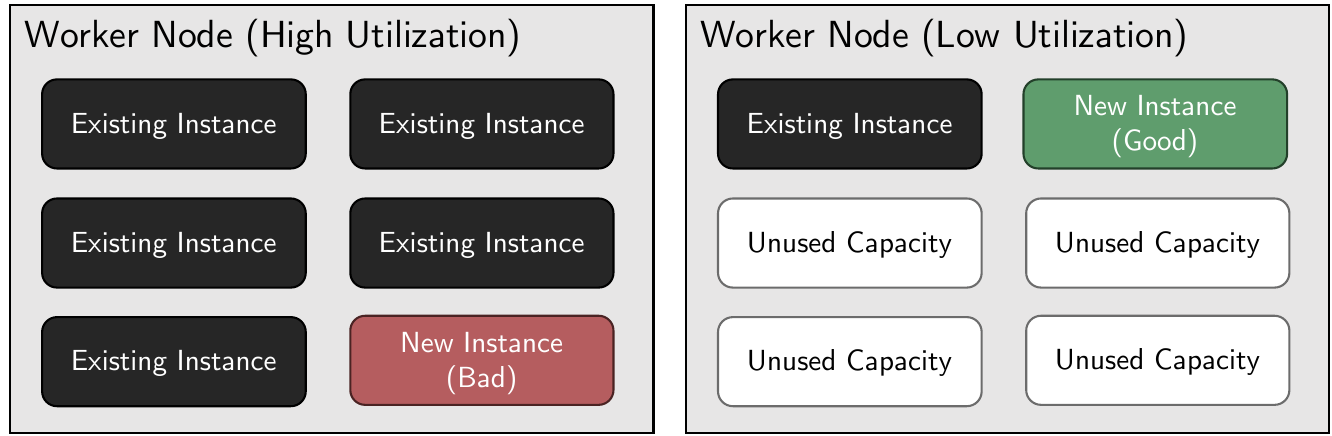}
    \caption{
        FaaS platforms start new function instances on shared worker nodes.
        While users can not influence which worker node will be used, underutilized nodes offer two benefits: faster execution times and lower cost.
        Using the \name{} system, newly started instances check if they are running on a node with low utilization.
        If not, they terminate themselves by crashing.
        This leads to a pool of better-performing instances that are re-used for subsequent invocations, leading to a compound performance increase.
    }
    \label{img:overview}
\end{figure}

Function-as-a-Service (FaaS) offerings have gained increasing popularity in the last years due to their ease of use~\cite{Manner_2023_Review,bermbach2017book}.
With FaaS, the cloud platform is responsible for handling most operational tasks, leaving users free to just write the code that should be executed for every incoming event, commonly referred to as functions~\cite{Schleier_2019_Berkeley,Manner_2023_Review}.
These functions can then be triggered by events in the cloud platform (such as a document upload) or by common transport protocols such as HTTP.
In response to incoming events, the cloud platform assigns a function instance~\cite{Brooker_2023_OnDemand}.
If there is no idle already running instance, the platform needs to start a new instance.
This process, called a cold start, leads to increased latency for the request as a worker node needs to load the user code and start an isolated environment to run the user code in~\cite{Manner_2018_ColdStarts}.
Often, platforms use microVMs~\cite{Brooker_2021_microVM} or containerization technologies for this~\cite{website_gvisorGCF}.
Users are billed based on the execution time of the invocation, with millisecond accuracy~\cite{Manner_2023_Review}.
This makes the model cost-efficient for use cases with irregular or infrequent usage patterns, since there is no resource overhead that users need to maintain.
As a cloud service hosted on shared infrastructure, functions suffer from performance variations between instances based on the usage patterns of the rest of the platform~\cite{bermbach2017book}.
For example, as shown in \cref{img:overview}, the worker node that the platform selects to start the instance can have different levels of overall utilization.
The instance on the worker node with low utilization will benefit from having to share its resources with fewer neighbors.
Taking the CPU as example this means that it will have to perform fewer context switches and that it is more likely data can remain cached.
These variations also concern most other cloud services.
However, FaaS users suffer twice when the variation leads to them having to use slower resources: not only is the execution time longer, but users are also billed \emph{more} for these slower requests due to the pay-per-use model.
In previous work~\cite{schirmer2023nightshift}, we have shown that some FaaS platforms have more than 10\% faster performance during the night, when the overall platform is under less load.
In this work, we focus on another kind of variation: when starting multiple instances in parallel, some of these instances are going to be faster than other ones.
Selectively only running invocations on the fast instances leads to faster \emph{and} cheaper function executions.
To enable this, we present \name{}, a system that runs a quick benchmark during every cold start.
If the benchmark reveals a particularly slow instance, \name{} re-queues the invocation and then terminates the function instance by crashing it.
This leads to the platform starting a new instance that might have better performance, or scheduling the invocation on an already running instance.
All instances that are already running must have passed the benchmark during their first request (as otherwise, they would have been shut down).
This leads to a pool of known-good instances that outperform the average instance, which is especially useful for longer running applications that re-use these instances often.
Our evaluation shows that this leads to a state where it is cheaper and faster for users to waste more resources of the platform.

In this paper, we make the following contributions:

\begin{itemize}
    \item We propose \name{}, a system to take advantage of these variations to select faster instances.
    \item We investigate how to choose which functions to keep and which should be terminated.
    \item We implement a proof-of-concept prototype for the FaaS platform Google Cloud Functions.
    \item We evaluate performance improvements for a data processing use case.
\end{itemize}

\begin{figure*}[!h]
    \centering
    \includegraphics[width=\textwidth]{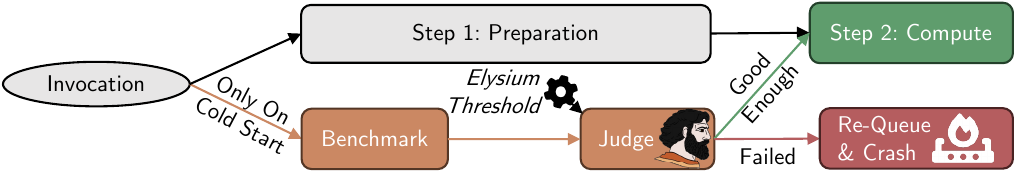}
    \caption{
        Overview of the process inside a function instance.
        The first step of the invocation, called prepare, is always executed.
        If the invocation is a cold start, a benchmark is executed in parallel.
        \name{} then judges whether the result is better than the \ef{}, in which case nothing happens (and the instance will be re-used in the future).
        If the instance does not pass, further execution is interrupted, the invocation is re-queued, and the instance is terminated.
    }
    \label{img:inside}
\end{figure*}

\section{A System to Exploit Platform Variability}

With \name{},\!\footnote{The system is named after the king from Greek mythology, who weights the souls of the dead to decide whether they should ascend to Elysium and enjoy eternal bliss or be thrown into the depths of Tartarus.}
users put their invocations into a queue, which triggers the platform to execute it.
The platform then places it in one of three kinds of execution environments:
Either it chooses an already warm instance, which is the ultimate goal of \name{}, as the instance has already been identified as having good performance.
The platform could also place the invocation on a new instance.
In this case, a quick benchmark is performed to measure a metric relevant to the later use case.
If the instance passes the benchmark, it will become a known-good instance and be re-used.
If not, it is terminated.
In order to not lose requests, before terminating, the instance re-queues the invocation that triggered it.

The following sections go into more detail on choosing how much should be terminated (\cref{subsec:howmuch}), how to decide on terminating in a distributed manner (\cref{subsec:elysium}), and how the benchmark should work (\cref{subsec:howlong}).

\subsection{How much to terminate?}\label{subsec:howmuch}

There is a trade-off in how good the performance in the benchmarking step needs to be so that the instance is not terminated.
Setting the required performance higher will lead to faster completion times per subsequent request, but it will also lead to many terminated (and subsequently re-queued) invocations, wasting resources.
Lower performance requirements lead to shorter completion times and cost in the short term, as requests are not re-queued as often.
However, over the long run, the lower average performance of the instances leads to longer completion times per request and thus higher cost.
Thus, the optimal termination rate depends on the duration of the workload, the performance variability of the platform, and the relative time (and thus cost) of the benchmark.
\begin{figure}
    \begin{equation*}
        \begin{split}
            c_{\textsf{total}} = & c_{\textsf{exec}} \left( \sum_{i=0}^{n_{\textsf{term}}} d_{\textsf{term}}+ \sum_{i=0}^{n_{\textsf{pass}}} d_{\textsf{pass}}+ \sum_{i=0}^{n_{\textsf{reuse}}} d_{\textsf{reuse}} \right) \\
                                 & + \underbrace{c_{\textsf{inv}} \left( n_{\textsf{term}} + n_{\textsf{pass}} + n_{\textsf{reuse}}\right)}_{\approx 0}
        \end{split}
    \end{equation*}
    \caption{The total cost of a workflow comprises the cost per millisecond of execution ($c_{\textsf{exec}}$) of all terminated ($d_{\textsf{term}}$), successful ($d_{\textsf{pass}}$) benchmarks and the re-used instances ($d_{\textsf{reuse}}$), as well as the cost per invocation ($c_{\textsf{inv}}$) per terminated, passed, and re-used function.
    }
    \label{fig:eq:cost}
\end{figure}

As shown in~\cref{fig:eq:cost}, the total cost of a workflow is influenced by the cost per unit of execution and the cost per invocation.
In general, the cost per unit of execution dominates the total cost:
For the smallest Google Cloud Functions function with 128 MB Memory, the cost per invocation ($c_{\textsf{inv}}$) is roughly equivalent to 50ms of execution time ($c_{\textsf{exec}}$).
For the biggest function with 32 GB Memory, it is less than 3 ms.
Other platforms follow a similar pricing pattern.
Thus, for all longer-running functions, the additional invocations made by \name{} are quickly offset by using faster instances.

As a kind of emergency exit, \name{} needs to keep track of how often an invocation has been re-queued already.
If an invocation has already caused to too many instance terminations compared to the expected rate, this means the platform performance is unusually slow or \name{} is unlucky with the instances it gets from the platform right now.
In this case, the function is marked as good without performing the benchmark, preventing infinite loops.
For example, if the expected termination rate is $40\%$, there is a ca. $1\%$ chance that an invocation fails five times in a row ($0.4^{5}$) and a less than 1\textperthousand{} chance that the same invocation experiences 8 terminations in a row.

\subsection{Autonomous Termination with the \EF{}}\label{subsec:elysium}
It would be optimal to have a centralized component that receives the benchmarking results from all instances to decide what instances are routed requests.
While this would allow for complicated scheduling mechanisms, it is not feasible for the kind of highly scalable and elastic workloads that FaaS platforms enable users to run.
Instead, a newly started function instance decides locally whether it is good enough based on just one value: what result in the benchmark is necessary to not terminate itself, i.e., so that the function will continue living and working on the next invocations.
We call this value the \emph{elysium factor}.
This is stored as part of the function configuration, so that \name{} does not require any outside communication during calls.

\paragraph{Calculating the \ef{} with pre-testing}

To calculate the \ef{}, \name{} runs a few benchmarking functions before executing the main workload.
This needs to be short enough to not influence the total duration and cost of the overall workflow significantly and may just comprise the first parts of the overall workload running without \name{} terminating instances.
Based on the performance values, \name{} can calculate the \ef{} to achieve a cost-optimal termination rate and then execute the rest of the workload with \name{} terminating instances.
As long as the whole workload runs short enough so that longer-term performance variations are not a significant influence (i.e., on the scale of hours), this is a simple and cost-effective approximation of the optimal termination rate.

\subsection{What and how long to benchmark}\label{subsec:howlong}

The kind of benchmark that should be performed and the acceptable \ef{} heavily depends on the use case.
Previous research has shown that FaaS workloads are most often constrained by network or CPU resources~\cite{Eismann_2021_Review}.
In our preliminary experiments, CPU performance showed the highest variability and was the bottleneck in most applications.

The benchmark itself should ideally not interfere with the normal execution of the function, i.e., it should run while the resource being benchmarked is otherwise unused.
The general approach is shown in~\cref{img:inside}.
During a first step of the execution, when resources need to be set up and downloaded, the benchmark and subsequent judging is run in parallel.
For example, in a data processing workflow, the raw data needs to be downloaded and loaded into memory.
At the same time, \name{} can run its benchmark.

On the other hand, if the function starts with a resource intensive operation, this first operation can be used as benchmark as well.
If the kind of operation at the start is representative for the rest of the function, this wastes fewer resources.

Overall, there is not one-size-fits-all solution for what kind of benchmark needs to be executed, and when as well as how long it should run.
\section{Evaluation}\label{sec:eval}

To demonstrate the feasibility of \name{} and evaluate potential performance gains we implement a proof of concept prototype and use it to run a data processing workload.

\subsection{Environment Configuration}
\subsubsection*{Use Case}

The use case in this evaluation is using linear regression analysis to predict weather.
To achieve this, the function first downloads a CSV file containing weather data for a specific location from previous days.
Afterwards, the data is analyzed to predict tomorrow's weather.
While downloading the file, \name{} can run its CPU benchmark as the download is network-limited.
The following linear regression analysis is CPU intensive, benefitting from the instance selection performed in the previous step.
To measure the capability of the CPU, we use matrix multiplication as benchmark in \name{}~\cite{Werner_2018_MatrixMult}.

\subsubsection*{Implementation}

We implement the prototype to work on Google Cloud Functions.
It does not use any features unique to the platform, so that the approach would work on different platforms as well.
The function is implemented in Go, a frequently used programming language for serverless applications.
The results for the evaluation are read from the function logs after the experiment is finished to rule out influences on execution duration.

\subsubsection*{Workload}

One experiment comprises ten virtual users (VU) sending a request and then waiting one second before sending the next request for 30 minutes.
We repeat this experiment every day at the same time for one week.\!\footnote{From 2025-02-03 to 2025-02-09, between 3 and 4 pm UTC}
As a baseline, we run the same workload at the same time using another function that is exactly the same, except that \name{} is completely disabled.
All experiments were executed in the europe-west-3 region of Google Cloud Platform with functions configured to have 256 MB RAM (leading to 0.167 vCPU).

Before these runs, we measured the \ef{} using pre-testing with a short running script where 10 VUs run for one minute (waiting one second between requests).
The \ef{} for the experiment is then set to the 60\textsuperscript{th} percentile of performance we measured (i.e., only the fastest 40\% of functions should pass the \ef{}), which is passed as parameter to the function by the VUs.

For both the baseline and experiment conditions, we measure the time of the function execution, the duration of the benchmark execution, whether it was successful, how often the invocation has been re-tried (retry count), how long the downloading step took (download duration), and how long the linear regression took (analysis duration).

\subsection{Results}

\subsubsection*{Linear Regression Duration}
\begin{figure}
    \centering
    \includegraphics[width=\linewidth]{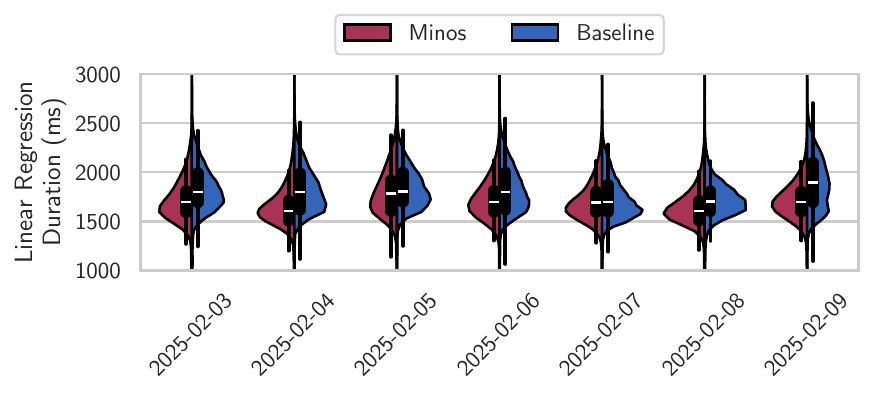}
    \caption{
        The linear regression step was, on average, faster every day the experiment was repeated.
        The maximum improvement was $>13\%$ on day two, and minimum improvement $4.3\%$ on day three and five.
        Note that the y-axis is limited to values between 1,000 and 3,000.
    }
    \label{img:eval:dayLr}
\end{figure}

The results for the average duration of the computationally heavy linear regression step are shown in~\cref{img:eval:dayLr}.
\name{} lead to better linear regression times on all days the experiment was repeated, with different effect sizes.
Over all days, \name{} improved the average regression duration by $7.8\%$.
This shows that \name{} reaches its core goal, reducing the duration of the resource intensive computation.

\subsubsection*{Successful Requests}

\begin{figure}
    \centering
    \includegraphics[width=\linewidth]{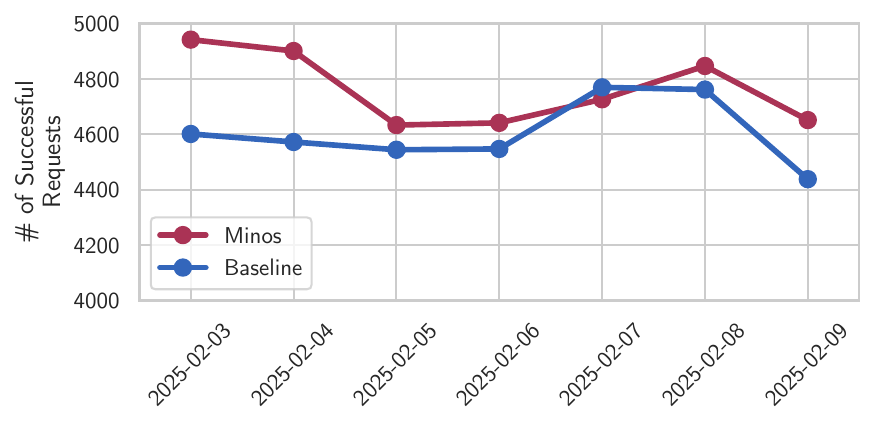}
    \caption{
        Successful requests per day.
        \name{} improved the amount of successful requests all days except one with a maximum of $7.3\%$ on day one, but reduced the amount by $<1\%$ on day five.
        Note that the y-axis is limited to values between 4,000 and 5,000.
    }
    \label{img:eval:daySucc}
\end{figure}

Extending the view from just looking at the resource intensive part of the function,~\cref{img:eval:daySucc} shows how many functions are successfully executed by the ten VUs each day.
On the first day, \name{} manages to complete 7.3\% more requests in the same 30 minutes.
This drops in the next days, and at day five, the baseline completed slightly more requests.
Over all experiments, \name{} completed 2.3\% more requests.

\subsubsection*{Cost}

\begin{figure}
    \centering
    \includegraphics[width=\linewidth]{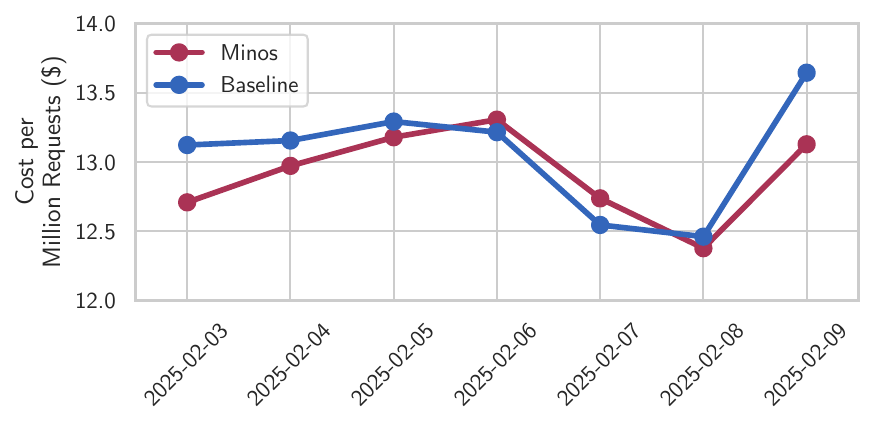}
    \caption{
        Average Cost per million successful requests per day the experiment ran.
        On the first and last day, \name{} saves more than $3\%$ cost, while it closely tracks the baseline on other days.
        Note that the y-axis is limited to values between 12 and 14.
    }
    \label{img:eval:dayCost}
\end{figure}

We calculate the cost per invocation based on the cost model of Google Cloud Functions.
For \name{}, we also take the invocations that ultimately were terminated into account and calculate the average cost per actually completed request (\emph{successful} request).
As shown in~\cref{img:eval:dayCost}, on day one, \name{} saves 3.3\% of cost, while also having 3.4\% lower latency.
This small difference is due to the cost per invocation.
On the other days, the cost per successful invocation of \name{} is very close to the baseline.
This shows that the platform itself is highly variable (as the experiment parameters never changed), and that the performance of \name{} depends on the overall cloud platform performance.
Over all experiments, \name{} was able to reduce the cost per request by 0.9\%.

\begin{figure}
    \centering
    \includegraphics[width=\linewidth]{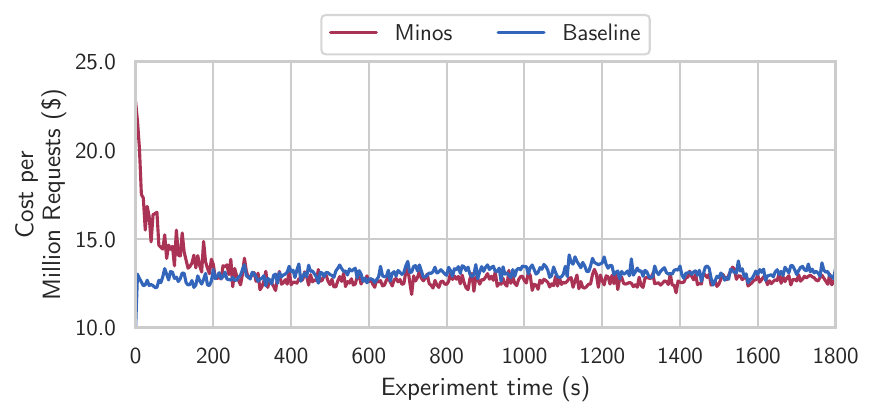}
    \caption{
        Average cost per million successful requests over all experiments.
        While the difference is small, \name{} has lower cost per successful request for 76\% of the experiment duration.
        Note that the y-axis is limited to values between 10 and 25.
    }
    \label{img:eval:expCost}
\end{figure}

Looking at how the cost develops over the duration of the experiments (cf. \cref{img:eval:expCost}) shows that, at the beginning, \name{} has a higher cost per successful request for the first 200s.
This is expected, as \name{} terminated many instances, increasing overall cost.
Afterwards, the average cost per successful invocation stays quite close, with \name{} staying slightly under the baseline for almost the whole duration.
After 670 s (11 min 10 s), \name{} was cheaper for more than 50 \% of time, which increases to 76\% at the end of the experiment.
This shows that letting \name{} run for a longer time increases its benefits.
\section{Discussion \& Future Work}

In the evaluation, \name{} showed a small but still considerable improvement over the baseline.
Using a data processing application, \name{} was able to handle up to 7.3\% more request and save up to 3\% cost while using \emph{more} overall resources from the computing platform.
This shows that the billing model of Google Cloud Functions, combined with its performance variability, motivates performance and cost oriented users to use more resources of the platform to achieve overall lower cost.
With a sufficiently large application, even relatively small savings still lead to considerable cost savings.
This poses a problem for the platforms, as their pricing model entices users to save money by wasting platform resources.
In general, while the goal of FaaS platforms is to abstract away as much operational complexity as possible, these kinds of applications demonstrate that users still benefit from understanding the platform more deeply.

\paragraph*{Cost Models}
Future work could focus on researching other cost models that make wasting resources of the cloud provider cost prohibitive.
One method could be billing users for the actual CPU performance delivered, not just the theoretical performance, to make less powerful instances financially applicable.
This would however require platforms to make it explicit that users can not control the exact pricing and speed of their invocations.
This is true today, it is just not represented in the cost model accordingly.
Another method would be for platforms to switch to fairer virtualization technologies.
For example, the popular FaaS platform AWS Lambda uses microVMs~\cite{Brooker_2021_microVM}, which in our preliminary experiments had a far fairer resource distribution with less overall variability.

\paragraph*{Online calculation of the \ef{}}
Our prototype for \name{} pre-calculates the \ef{}.
It could also be updated live based on the benchmarking results of currently running instances.
After a benchmark has finished, instances can report their results to a centralized component that collects live performance data.
It then periodically re-calculates the \ef{} and updates the functions comparable to our approach in our previous work~\cite{schirmer2024fusionizepp}.
While this component is centralized, it is not a single point of failure as its failures would just lead to temporarily suboptimal performance.
Depending on the scale of the application, it is not feasible to store all past benchmark results to accurately calculate the performance percentiles.
Instead, some statistical properties can be calculated online, i.e., only having to store the most recent new value and a limited amount of information about past results.
For example, the mean can be updated online by storing the previous mean and the amount of previous results.
Other values, such as the standard deviation~\cite{Jain_1985_Psquared} or percentiles~\cite{Welford_1962_algo} can only be estimated online and not calculated exactly.

\paragraph*{Workload Limitations}
Note that \name{} requires an asynchronous workload, i.e., users need to put their invocations into a queue instead of synchronously waiting for a response.
When users require a synchronous response, the instance that receives the request needs to stay alive until the computation is finished to send the response back.
If the instance fails the benchmark and should be crashed, it would need to send the call to another function and wait for the response, leading to double billing as the waiting instance still incurs cost~\cite{Baldini_2017_Trilemma}.
\section{Related Work}
% PERFORMANCE VARIATIONS
Overall variability of FaaS functions has, e.g., been shown over longer~\cite{Eismann_2022_LongTerm, Ustiugov_2021_Stellar} and shorter~\cite{schirmer2023profaastinate} time frames, between processor architectures~\cite{Lambion_2022_ArmConsistency}, between platforms~\cite{grambow2024befaas_tr}, and for the overall platform behavior~\cite{Mahmoudi_2021_FaaSPlatformPerf,schirmer2023nightshift,Ustiugov_2021_Stellar}.
Besides compute performance, instance placement can also be used to optimize for fewer carbon emissions~\cite{Chadha_2023_GreenCourier,Cordingly_2024_SkyComputing}.

Ginzburg and Freedman present a method for using short-term intra-region FaaS instance variability on AWS Lambda~\cite{Ginzburg_2021_Variability}.
They demonstrate temporal, spatial (i.e., using other cloud regions), and instance performance variations in AWS Lambda in 2021.
Their evaluation shows that exploiting instance performance variability can lead to performance increases.
Compared to \name{}, their approach focuses on demonstrating the existence of the effect for CPUs and relies on a centralized scheduler, only working for a limited amount of instances.
% Die deployen die gleiche Funktion 50x, starten dann davon jeweils eine Instanz, und machen dann centralized scheudling zu den 60% schnellsten.

% COLD STARTS
Another way of optimizing FaaS applications is speeding up cold starts, e.g., by pre-warming functions before they are called~\cite{bermbach2020coldstarts, Bardsley_2018_ColdStartPrewarm,carl2024geoff,Daw_2020_Xandau}.
This can be combined with \name{} by benchmarking the pre-warmed instances before they are used.
Other approaches focus on changing the underlying platform to speed up cold starts, e.g., by changing the virtualization and storage layer~\cite{Brooker_2023_OnDemand}, or by switching out the scheduling mechanism ~\cite{Cvetkovic_2024_Dirigent}.

% OFFLOADING FROM FAAS
Another approach is only using FaaS for the parts of the application where its most useful.
Other parts can be executed to less expensive compute services~\cite{Zhao_2023_BeeHive,Elsakhawy_2023_UnFaaSener}.
\section{Conclusion}

In this paper we have presented \name{}, a system to take advantage of FaaS platform variability to save cost and decrease latency.
The system benchmarks newly started instances and deliberately terminates them if they do not perform fast enough.
This leads to only performant instances being re-used for later invocations, over time leading to a faster pool of instances.
Using a prototype built on Google Cloud Functions and a data processing use case, we were able to demonstrate cost savings of up to 3.2\% per invocation while handling 7.3\% more requests in the same time.
Our research shows that cloud platforms need to take precaution to disable these kinds of use cases, as users can waste more platform resources while at the same time paying less overall.
Future research can focus on developing fairer virtualization technologies or researching novel cost models.

\begin{acks}
    Partially funded by the \grantsponsor{BMBF}{Bundesministerium für Bildung und Forschung (BMBF, German Federal Ministry of Education and Research)}{https://www.bmbf.de/bmbf/en} in the scope of the Software Campus 3.0 (Technische Universit\"at Berlin) program -- \grantnum{BMBF}{01IS23068}.
\end{acks}

\balance

\bibliographystyle{ACM-Reference-Format}
\bibliography{bibliography.bib}

%%% -*-BibTeX-*-
%%% Do NOT edit. File created by BibTeX with style
%%% ACM-Reference-Format-Journals [18-Jan-2012].

\begin{thebibliography}{30}

%%% ====================================================================
%%% NOTE TO THE USER: you can override these defaults by providing
%%% customized versions of any of these macros before the \bibliography
%%% command.  Each of them MUST provide its own final punctuation,
%%% except for \shownote{} and \showURL{}.  The latter two
%%% do not use final punctuation, in order to avoid confusing it with
%%% the Web address.
%%%
%%% To suppress output of a particular field, define its macro to expand
%%% to an empty string, or better, \unskip, like this:
%%%
%%% \newcommand{\showURL}[1]{\unskip}   % LaTeX syntax
%%%
%%% \def \showURL #1{\unskip}           % plain TeX syntax
%%%
%%% ====================================================================

\ifx \showCODEN    \undefined \def \showCODEN     #1{\unskip}     \fi
\ifx \showISBNx    \undefined \def \showISBNx     #1{\unskip}     \fi
\ifx \showISBNxiii \undefined \def \showISBNxiii  #1{\unskip}     \fi
\ifx \showISSN     \undefined \def \showISSN      #1{\unskip}     \fi
\ifx \showLCCN     \undefined \def \showLCCN      #1{\unskip}     \fi
\ifx \shownote     \undefined \def \shownote      #1{#1}          \fi
\ifx \showarticletitle \undefined \def \showarticletitle #1{#1}   \fi
\ifx \showURL      \undefined \def \showURL       {\relax}        \fi
% The following commands are used for tagged output and should be
% invisible to TeX
\providecommand\bibfield[2]{#2}
\providecommand\bibinfo[2]{#2}
\providecommand\natexlab[1]{#1}
\providecommand\showeprint[2][]{arXiv:#2}

\bibitem[and(1962)]%
        {Welford_1962_algo}
\bibfield{author}{\bibinfo{person}{B.~P.~Welford and}.} \bibinfo{year}{1962}\natexlab{}.
\newblock \showarticletitle{Note on a Method for Calculating Corrected Sums of Squares and Products}.
\newblock \bibinfo{journal}{\emph{Technometrics}} \bibinfo{volume}{4}, \bibinfo{number}{3} (\bibinfo{year}{1962}), \bibinfo{pages}{419--420}.
\newblock
\href{https://doi.org/10.1080/00401706.1962.10490022}{doi:\nolinkurl{10.1080/00401706.1962.10490022}}
\showeprint{https://www.tandfonline.com/doi/pdf/10.1080/00401706.1962.10490022}


\bibitem[Baldini et~al\mbox{.}(2017)]%
        {Baldini_2017_Trilemma}
\bibfield{author}{\bibinfo{person}{Ioana Baldini}, \bibinfo{person}{Perry Cheng}, \bibinfo{person}{Stephen~J. Fink}, \bibinfo{person}{Nick Mitchell}, \bibinfo{person}{Vinod Muthusamy}, \bibinfo{person}{Rodric Rabbah}, \bibinfo{person}{Philippe Suter}, {and} \bibinfo{person}{Olivier Tardieu}.} \bibinfo{year}{2017}\natexlab{}.
\newblock \showarticletitle{The serverless trilemma: function composition for serverless computing}. In \bibinfo{booktitle}{\emph{Proceedings of the 2017 ACM SIGPLAN International Symposium on New Ideas, New Paradigms, and Reflections on Programming and Software}}. \bibinfo{publisher}{ACM}, \bibinfo{address}{Vancouver BC Canada}, \bibinfo{pages}{89–103}.
\newblock
\showISBNx{978-1-4503-5530-8}
\href{https://doi.org/10.1145/3133850.3133855}{doi:\nolinkurl{10.1145/3133850.3133855}}


\bibitem[Bardsley et~al\mbox{.}(2018)]%
        {Bardsley_2018_ColdStartPrewarm}
\bibfield{author}{\bibinfo{person}{Daniel Bardsley}, \bibinfo{person}{Larry Ryan}, {and} \bibinfo{person}{John Howard}.} \bibinfo{year}{2018}\natexlab{}.
\newblock \showarticletitle{Serverless Performance and Optimization Strategies}. In \bibinfo{booktitle}{\emph{2018 IEEE International Conference on Smart Cloud (SmartCloud)}}. \bibinfo{pages}{19–26}.
\newblock
\href{https://doi.org/10.1109/SmartCloud.2018.00012}{doi:\nolinkurl{10.1109/SmartCloud.2018.00012}}


\bibitem[Bermbach et~al\mbox{.}(2020)]%
        {bermbach2020coldstarts}
\bibfield{author}{\bibinfo{person}{David Bermbach}, \bibinfo{person}{Ahmet-Serdar Karakaya}, {and} \bibinfo{person}{Simon Buchholz}.} \bibinfo{year}{2020}\natexlab{}.
\newblock \showarticletitle{Using Application Knowledge to Reduce Cold Starts in FaaS Services}. In \bibinfo{booktitle}{\emph{Proceedings of the 35th ACM Symposium on Applied Computing}} (Brno, Czech Republic) \emph{(\bibinfo{series}{SAC '20})}. \bibinfo{publisher}{Association for Computing Machinery (ACM)}, \bibinfo{address}{New York, NY, USA}, \bibinfo{pages}{134--143}.
\newblock
\href{https://doi.org/10.1145/3341105.3373909}{doi:\nolinkurl{10.1145/3341105.3373909}}


\bibitem[Bermbach et~al\mbox{.}(2017)]%
        {bermbach2017book}
\bibfield{author}{\bibinfo{person}{David Bermbach}, \bibinfo{person}{Erik Wittern}, {and} \bibinfo{person}{Stefan Tai}.} \bibinfo{year}{2017}\natexlab{}.
\newblock \bibinfo{booktitle}{\emph{Cloud Service Benchmarking: Measuring Quality of Cloud Services from a Client Perspective}}.
\newblock \bibinfo{publisher}{Springer}, \bibinfo{address}{Cham, Switzerland}.
\newblock
\showISBNx{978-3-319-85672-8}


\bibitem[Brooker et~al\mbox{.}(2021)]%
        {Brooker_2021_microVM}
\bibfield{author}{\bibinfo{person}{Marc Brooker}, \bibinfo{person}{Adrian~Costin Catangiu}, \bibinfo{person}{Mike Danilov}, \bibinfo{person}{Alexander Graf}, \bibinfo{person}{Colm MacCarthaigh}, {and} \bibinfo{person}{Andrei Sandu}.} \bibinfo{year}{2021}\natexlab{}.
\newblock \showarticletitle{Restoring Uniqueness in MicroVM Snapshots}.
\newblock  \bibinfo{number}{arXiv:2102.12892} (\bibinfo{date}{Feb.} \bibinfo{year}{2021}).
\newblock
\urldef\tempurl%
\url{http://arxiv.org/abs/2102.12892}
\showURL{%
\tempurl}
\newblock
\shownote{arXiv:2102.12892 [cs]}.


\bibitem[Brooker et~al\mbox{.}(2023)]%
        {Brooker_2023_OnDemand}
\bibfield{author}{\bibinfo{person}{Marc Brooker}, \bibinfo{person}{Mike Danilov}, \bibinfo{person}{Chris Greenwood}, {and} \bibinfo{person}{Phil Piwonka}.} \bibinfo{year}{2023}\natexlab{}.
\newblock \showarticletitle{On-demand Container Loading in {AWS} Lambda}. \bibinfo{pages}{315–328}.
\newblock
\showISBNx{978-1-939133-35-9}
\urldef\tempurl%
\url{https://www.usenix.org/conference/atc23/presentation/brooker}
\showURL{%
\tempurl}


\bibitem[Carl et~al\mbox{.}(2024)]%
        {carl2024geoff}
\bibfield{author}{\bibinfo{person}{Natalie Carl}, \bibinfo{person}{Trever Schirmer}, \bibinfo{person}{Tobias Pfandzelter}, {and} \bibinfo{person}{David Bermbach}.} \bibinfo{year}{2024}\natexlab{}.
\newblock \showarticletitle{GeoFF: Federated Serverless Workflows with Data Pre-Fetching}. In \bibinfo{booktitle}{\emph{Proceedings of the 12th IEEE International Conference on Cloud Engineering}} (Paphos, Cyprus) \emph{(\bibinfo{series}{IC2E '24})}. \bibinfo{publisher}{IEEE}, \bibinfo{address}{New York, NY, USA}, \bibinfo{pages}{144--151}.
\newblock
\href{https://doi.org/10.1109/IC2E61754.2024.00023}{doi:\nolinkurl{10.1109/IC2E61754.2024.00023}}


\bibitem[Chadha et~al\mbox{.}(2023)]%
        {Chadha_2023_GreenCourier}
\bibfield{author}{\bibinfo{person}{Mohak Chadha}, \bibinfo{person}{Thandayuthapani Subramanian}, \bibinfo{person}{Eishi Arima}, \bibinfo{person}{Michael Gerndt}, \bibinfo{person}{Martin Schulz}, {and} \bibinfo{person}{Osama Abboud}.} \bibinfo{year}{2023}\natexlab{}.
\newblock \showarticletitle{GreenCourier: Carbon-Aware Scheduling for Serverless Functions}. In \bibinfo{booktitle}{\emph{Proceedings of the 9th International Workshop on Serverless Computing}} (Bologna, Italy) \emph{(\bibinfo{series}{WoSC '23})}. \bibinfo{publisher}{Association for Computing Machinery}, \bibinfo{address}{New York, NY, USA}, \bibinfo{pages}{18–23}.
\newblock
\showISBNx{9798400704550}
\href{https://doi.org/10.1145/3631295.3631396}{doi:\nolinkurl{10.1145/3631295.3631396}}


\bibitem[Cordingly et~al\mbox{.}(2023)]%
        {Cordingly_2024_SkyComputing}
\bibfield{author}{\bibinfo{person}{Robert Cordingly}, \bibinfo{person}{Jasleen Kaur}, \bibinfo{person}{Divyansh Dwivedi}, {and} \bibinfo{person}{Wes Lloyd}.} \bibinfo{year}{2023}\natexlab{}.
\newblock \showarticletitle{Towards Serverless Sky Computing: An Investigation on Global Workload Distribution to Mitigate Carbon Intensity, Network Latency, and Cost}. In \bibinfo{booktitle}{\emph{2023 IEEE International Conference on Cloud Engineering (IC2E)}}. \bibinfo{pages}{59--69}.
\newblock
\href{https://doi.org/10.1109/IC2E59103.2023.00015}{doi:\nolinkurl{10.1109/IC2E59103.2023.00015}}


\bibitem[Cvetković et~al\mbox{.}(2024)]%
        {Cvetkovic_2024_Dirigent}
\bibfield{author}{\bibinfo{person}{Lazar Cvetković}, \bibinfo{person}{François Costa}, \bibinfo{person}{Mihajlo Djokic}, \bibinfo{person}{Michal Friedman}, {and} \bibinfo{person}{Ana Klimovic}.} \bibinfo{year}{2024}\natexlab{}.
\newblock \showarticletitle{Dirigent: Lightweight Serverless Orchestration}. In \bibinfo{booktitle}{\emph{Proceedings of the ACM SIGOPS 30th Symposium on Operating Systems Principles}} \emph{(\bibinfo{series}{SOSP ’24})}. \bibinfo{publisher}{Association for Computing Machinery}, \bibinfo{address}{New York, NY, USA}, \bibinfo{pages}{369–384}.
\newblock
\showISBNx{9798400712517}
\href{https://doi.org/10.1145/3694715.3695966}{doi:\nolinkurl{10.1145/3694715.3695966}}


\bibitem[Daw et~al\mbox{.}(2020)]%
        {Daw_2020_Xandau}
\bibfield{author}{\bibinfo{person}{Nilanjan Daw}, \bibinfo{person}{Umesh Bellur}, {and} \bibinfo{person}{Purushottam Kulkarni}.} \bibinfo{year}{2020}\natexlab{}.
\newblock \showarticletitle{Xanadu: Mitigating cascading cold starts in serverless function chain deployments}. \bibinfo{pages}{356–370}.
\newblock
\href{https://doi.org/10.1145/3423211.3425690}{doi:\nolinkurl{10.1145/3423211.3425690}}


\bibitem[Eismann et~al\mbox{.}(2022)]%
        {Eismann_2022_LongTerm}
\bibfield{author}{\bibinfo{person}{Simon Eismann}, \bibinfo{person}{Diego~Elias Costa}, \bibinfo{person}{Lizhi Liao}, \bibinfo{person}{Cor-Paul Bezemer}, \bibinfo{person}{Weiyi Shang}, \bibinfo{person}{André van Hoorn}, {and} \bibinfo{person}{Samuel Kounev}.} \bibinfo{year}{2022}\natexlab{}.
\newblock \showarticletitle{A case study on the stability of performance tests for serverless applications}.
\newblock \bibinfo{journal}{\emph{Journal of Systems and Software}}  \bibinfo{volume}{189} (\bibinfo{date}{July} \bibinfo{year}{2022}), \bibinfo{pages}{111294}.
\newblock
\showISSN{0164-1212}
\href{https://doi.org/10.1016/j.jss.2022.111294}{doi:\nolinkurl{10.1016/j.jss.2022.111294}}


\bibitem[Eismann et~al\mbox{.}(2021)]%
        {Eismann_2021_Review}
\bibfield{author}{\bibinfo{person}{Simon Eismann}, \bibinfo{person}{Joel Scheuner}, \bibinfo{person}{Erwin van Eyk}, \bibinfo{person}{Maximilian Schwinger}, \bibinfo{person}{Johannes Grohmann}, \bibinfo{person}{Nikolas Herbst}, \bibinfo{person}{Cristina~L. Abad}, {and} \bibinfo{person}{Alexandru Iosup}.} \bibinfo{year}{2021}\natexlab{}.
\newblock \showarticletitle{A Review of Serverless Use Cases and their Characteristics}.
\newblock  \bibinfo{number}{arXiv:2008.11110} (\bibinfo{date}{Jan.} \bibinfo{year}{2021}).
\newblock
\href{https://doi.org/10.48550/arXiv.2008.11110}{doi:\nolinkurl{10.48550/arXiv.2008.11110}}
\newblock
\shownote{arXiv:2008.11110 [cs]}.


\bibitem[Ginzburg and Freedman(2021)]%
        {Ginzburg_2021_Variability}
\bibfield{author}{\bibinfo{person}{Samuel Ginzburg} {and} \bibinfo{person}{Michael~J. Freedman}.} \bibinfo{year}{2021}\natexlab{}.
\newblock \showarticletitle{Serverless Isn’t Server-Less: Measuring and Exploiting Resource Variability on Cloud FaaS Platforms}. In \bibinfo{booktitle}{\emph{Proceedings of the 2020 Sixth International Workshop on Serverless Computing}} \emph{(\bibinfo{series}{WoSC’20})}. \bibinfo{publisher}{Association for Computing Machinery}, \bibinfo{address}{New York, NY, USA}, \bibinfo{pages}{43–48}.
\newblock
\showISBNx{978-1-4503-8204-5}
\href{https://doi.org/10.1145/3429880.3430099}{doi:\nolinkurl{10.1145/3429880.3430099}}


\bibitem[Grambow et~al\mbox{.}(2024)]%
        {grambow2024befaas_tr}
\bibfield{author}{\bibinfo{person}{Martin Grambow}, \bibinfo{person}{Tobias Pfandzelter}, {and} \bibinfo{person}{David Bermbach}.} \bibinfo{year}{2024}\natexlab{}.
\newblock \bibinfo{booktitle}{\emph{Application-Centric Benchmarking of Distributed FaaS Platforms using BeFaaS}}.
\newblock \bibinfo{type}{{T}echnical {R}eport} 3S.2024.1. \bibinfo{institution}{Technische Universit{\"a}t Berlin, Scalable Software Systems Research Group}, \bibinfo{address}{Berlin, Germany}.
\newblock


\bibitem[Jain and Chlamtac(1985)]%
        {Jain_1985_Psquared}
\bibfield{author}{\bibinfo{person}{Raj Jain} {and} \bibinfo{person}{Imrich Chlamtac}.} \bibinfo{year}{1985}\natexlab{}.
\newblock \showarticletitle{The P2 algorithm for dynamic calculation of quantiles and histograms without storing observations}.
\newblock \bibinfo{journal}{\emph{Commun. ACM}} \bibinfo{volume}{28}, \bibinfo{number}{10} (\bibinfo{date}{Oct.} \bibinfo{year}{1985}), \bibinfo{pages}{1076--1085}.
\newblock
\showISSN{0001-0782}
\href{https://doi.org/10.1145/4372.4378}{doi:\nolinkurl{10.1145/4372.4378}}


\bibitem[Jonas et~al\mbox{.}(2019)]%
        {Schleier_2019_Berkeley}
\bibfield{author}{\bibinfo{person}{Eric Jonas}, \bibinfo{person}{Johann Schleier-Smith}, \bibinfo{person}{Vikram Sreekanti}, \bibinfo{person}{Chia-Che Tsai}, \bibinfo{person}{Anurag Khandelwal}, \bibinfo{person}{Qifan Pu}, \bibinfo{person}{Vaishaal Shankar}, \bibinfo{person}{Joao Carreira}, \bibinfo{person}{Karl Krauth}, \bibinfo{person}{Neeraja Yadwadkar}, \bibinfo{person}{Joseph~E. Gonzalez}, \bibinfo{person}{Raluca~Ada Popa}, \bibinfo{person}{Ion Stoica}, {and} \bibinfo{person}{David~A. Patterson}.} \bibinfo{year}{2019}\natexlab{}.
\newblock \showarticletitle{Cloud Programming Simplified: A Berkeley View on Serverless Computing}.
\newblock \bibinfo{journal}{\emph{arXiv:1902.03383 [cs]}} (\bibinfo{year}{2019}).
\newblock
\newblock
\shownote{arXiv: 1902.03383}.


\bibitem[Lambion et~al\mbox{.}(2022)]%
        {Lambion_2022_ArmConsistency}
\bibfield{author}{\bibinfo{person}{Danielle Lambion}, \bibinfo{person}{Robert Schmitz}, \bibinfo{person}{Robert Cordingly}, \bibinfo{person}{Navid Heydari}, {and} \bibinfo{person}{Wes Lloyd}.} \bibinfo{year}{2022}\natexlab{}.
\newblock \showarticletitle{Characterizing X86 and ARM Serverless Performance Variation: A Natural Language Processing Case Study}. In \bibinfo{booktitle}{\emph{Companion of the 2022 ACM/SPEC International Conference on Performance Engineering}}. \bibinfo{publisher}{ACM}, \bibinfo{address}{Bejing China}, \bibinfo{pages}{69–75}.
\newblock
\showISBNx{978-1-4503-9159-7}
\href{https://doi.org/10.1145/3491204.3543506}{doi:\nolinkurl{10.1145/3491204.3543506}}


\bibitem[Mahmoudi and Khazaei(2021)]%
        {Mahmoudi_2021_FaaSPlatformPerf}
\bibfield{author}{\bibinfo{person}{Nima Mahmoudi} {and} \bibinfo{person}{Hamzeh Khazaei}.} \bibinfo{year}{2021}\natexlab{}.
\newblock \showarticletitle{Temporal Performance Modelling of Serverless Computing Platforms}. In \bibinfo{booktitle}{\emph{Proceedings of the 2020 Sixth International Workshop on Serverless Computing}} \emph{(\bibinfo{series}{WoSC’20})}. \bibinfo{publisher}{Association for Computing Machinery}, \bibinfo{address}{New York, NY, USA}, \bibinfo{pages}{1–6}.
\newblock
\showISBNx{978-1-4503-8204-5}
\href{https://doi.org/10.1145/3429880.3430092}{doi:\nolinkurl{10.1145/3429880.3430092}}


\bibitem[Manner(2023)]%
        {Manner_2023_Review}
\bibfield{author}{\bibinfo{person}{Johannes Manner}.} \bibinfo{year}{2023}\natexlab{}.
\newblock \showarticletitle{A Structured Literature Review Approach to Define Serverless Computing and Function as a Service}. In \bibinfo{booktitle}{\emph{2023 IEEE 16th International Conference on Cloud Computing (CLOUD)}}. \bibinfo{pages}{516–522}.
\newblock
\showISSN{2159-6190}
\href{https://doi.org/10.1109/CLOUD60044.2023.00068}{doi:\nolinkurl{10.1109/CLOUD60044.2023.00068}}


\bibitem[Manner et~al\mbox{.}(2018)]%
        {Manner_2018_ColdStarts}
\bibfield{author}{\bibinfo{person}{Johannes Manner}, \bibinfo{person}{Martin EndreB}, \bibinfo{person}{Tobias Heckel}, {and} \bibinfo{person}{Guido Wirtz}.} \bibinfo{year}{2018}\natexlab{}.
\newblock \showarticletitle{Cold Start Influencing Factors in Function as a Service}. In \bibinfo{booktitle}{\emph{2018 IEEE/ACM International Conference on Utility and Cloud Computing Companion (UCC Companion)}}. \bibinfo{publisher}{IEEE}, \bibinfo{address}{Zurich}, \bibinfo{pages}{181–188}.
\newblock
\showISBNx{978-1-7281-0359-4}
\href{https://doi.org/10.1109/UCC-Companion.2018.00054}{doi:\nolinkurl{10.1109/UCC-Companion.2018.00054}}


\bibitem[Platfomr(2020)]%
        {website_gvisorGCF}
\bibfield{author}{\bibinfo{person}{Google~Cloud Platfomr}.} \bibinfo{year}{2020}\natexlab{}.
\newblock \bibinfo{booktitle}{\emph{How gVisor protects Gooogle Cloud Services from CVE-2020-14386}}.
\newblock
\urldef\tempurl%
\url{https://cloud.google.com/blog/products/containers-kubernetes/how-gvisor-protects-google-cloud-services-from-cve-2020-14386?hl=en}
\showURL{%
Retrieved 2024-10-24 from \tempurl}


\bibitem[Sadeghian et~al\mbox{.}(2023)]%
        {Elsakhawy_2023_UnFaaSener}
\bibfield{author}{\bibinfo{person}{Ghazal Sadeghian}, \bibinfo{person}{Mohamed Elsakhawy}, \bibinfo{person}{Mohanna Shahrad}, \bibinfo{person}{Joe Hattori}, {and} \bibinfo{person}{Mohammad Shahrad}.} \bibinfo{year}{2023}\natexlab{}.
\newblock \showarticletitle{{UnFaaSener}: Latency and Cost Aware Offloading of Functions from Serverless Platforms}. \bibinfo{pages}{879–896}.
\newblock
\showISBNx{978-1-939133-35-9}
\urldef\tempurl%
\url{https://www.usenix.org/conference/atc23/presentation/sadeghian}
\showURL{%
\tempurl}


\bibitem[Schirmer et~al\mbox{.}(2023a)]%
        {schirmer2023profaastinate}
\bibfield{author}{\bibinfo{person}{Trever Schirmer}, \bibinfo{person}{Natalie Carl}, \bibinfo{person}{Tobias Pfandzelter}, {and} \bibinfo{person}{David Bermbach}.} \bibinfo{year}{2023}\natexlab{a}.
\newblock \showarticletitle{ProFaaStinate: Delaying Serverless Function Calls to Optimize Platform Performance}. In \bibinfo{booktitle}{\emph{Proceedings of the 9th International Workshop on Serverless Computing}} (Bologna, Italy) \emph{(\bibinfo{series}{WoSC '23})}. \bibinfo{publisher}{Association for Computing Machinery (ACM)}, \bibinfo{address}{New York, NY, USA}, \bibinfo{pages}{1--6}.
\newblock
\href{https://doi.org/10.1145/3631295.3631393}{doi:\nolinkurl{10.1145/3631295.3631393}}


\bibitem[Schirmer et~al\mbox{.}(2023b)]%
        {schirmer2023nightshift}
\bibfield{author}{\bibinfo{person}{Trever Schirmer}, \bibinfo{person}{Nils Japke}, \bibinfo{person}{Sofia Greten}, \bibinfo{person}{Tobias Pfandzelter}, {and} \bibinfo{person}{David Bermbach}.} \bibinfo{year}{2023}\natexlab{b}.
\newblock \showarticletitle{The Night Shift: Understanding Performance Variability of Cloud Serverless Platforms}. In \bibinfo{booktitle}{\emph{Proceedings of the 1st Workshop on SErverless Systems, Applications and MEthodologies}} (Rome, Italy) \emph{(\bibinfo{series}{SESAME '23})}. \bibinfo{publisher}{Association for Computing Machinery (ACM)}, \bibinfo{address}{New York, NY, USA}.
\newblock
\href{https://doi.org/10.1145/3592533.3592808}{doi:\nolinkurl{10.1145/3592533.3592808}}


\bibitem[Schirmer et~al\mbox{.}(2024)]%
        {schirmer2024fusionizepp}
\bibfield{author}{\bibinfo{person}{Trever Schirmer}, \bibinfo{person}{Joel Scheuner}, \bibinfo{person}{Tobias Pfandzelter}, {and} \bibinfo{person}{David Bermbach}.} \bibinfo{year}{2024}\natexlab{}.
\newblock \showarticletitle{FUSIONIZE++: Improving Serverless Application Performance Using Dynamic Task Inlining and Infrastructure Optimization}.
\newblock \bibinfo{journal}{\emph{IEEE Transactions on Cloud Computing}} (\bibinfo{date}{Aug.} \bibinfo{year}{2024}).
\newblock
\showISSN{2168-7161}
\href{https://doi.org/10.1109/TCC.2024.3451108}{doi:\nolinkurl{10.1109/TCC.2024.3451108}}


\bibitem[Ustiugov et~al\mbox{.}(2021)]%
        {Ustiugov_2021_Stellar}
\bibfield{author}{\bibinfo{person}{Dmitrii Ustiugov}, \bibinfo{person}{Theodor Amariucai}, {and} \bibinfo{person}{Boris Grot}.} \bibinfo{year}{2021}\natexlab{}.
\newblock \showarticletitle{Analyzing Tail Latency in Serverless Clouds with STeLLAR}. In \bibinfo{booktitle}{\emph{2021 IEEE International Symposium on Workload Characterization (IISWC)}}. \bibinfo{pages}{51–62}.
\newblock
\href{https://doi.org/10.1109/IISWC53511.2021.00016}{doi:\nolinkurl{10.1109/IISWC53511.2021.00016}}


\bibitem[Werner et~al\mbox{.}(2018)]%
        {Werner_2018_MatrixMult}
\bibfield{author}{\bibinfo{person}{Sebastian Werner}, \bibinfo{person}{Jörn Kuhlenkamp}, \bibinfo{person}{Markus Klems}, \bibinfo{person}{Johannes Müller}, {and} \bibinfo{person}{Stefan Tai}.} \bibinfo{year}{2018}\natexlab{}.
\newblock \showarticletitle{Serverless Big Data Processing using Matrix Multiplication as Example}. In \bibinfo{booktitle}{\emph{2018 IEEE International Conference on Big Data (Big Data)}}. \bibinfo{pages}{358--365}.
\newblock
\href{https://doi.org/10.1109/BigData.2018.8622362}{doi:\nolinkurl{10.1109/BigData.2018.8622362}}


\bibitem[Zhao et~al\mbox{.}(2023)]%
        {Zhao_2023_BeeHive}
\bibfield{author}{\bibinfo{person}{Ziming Zhao}, \bibinfo{person}{Mingyu Wu}, \bibinfo{person}{Jiawei Tang}, \bibinfo{person}{Binyu Zang}, \bibinfo{person}{Zhaoguo Wang}, {and} \bibinfo{person}{Haibo Chen}.} \bibinfo{year}{2023}\natexlab{}.
\newblock \showarticletitle{BeeHive: Sub-second Elasticity for Web Services with Semi-FaaS Execution}. In \bibinfo{booktitle}{\emph{Proceedings of the 28th ACM International Conference on Architectural Support for Programming Languages and Operating Systems, Volume 2}}. \bibinfo{publisher}{ACM}, \bibinfo{address}{Vancouver BC Canada}, \bibinfo{pages}{74–87}.
\newblock
\showISBNx{978-1-4503-9916-6}
\href{https://doi.org/10.1145/3575693.3575752}{doi:\nolinkurl{10.1145/3575693.3575752}}


\end{thebibliography}

\end{document}